\documentstyle[preprint,revtex]{aps}
\begin{document}
%\draft
\preprint{January 20, 1993}
\begin{title}
Universal magnetic properties of $La_{2-\delta} Sr_{\delta} CuO_{4}$ \\
at intermediate temperatures.
\end{title}
\author{Andrey V. Chubukov${}^{1,2}$ and Subir Sachdev${}^{1}$}
\begin{instit}
${}^{1}$Departments of Physics and Applied Physics, P.O. Box 2157,\\
Yale University, New Haven, CT 06520,\\
and ${}^{2}$P.L. Kapitza Institute for Physical Problems, Moscow, Russia
\end{instit}
\begin{abstract}
We present the theory of two-dimensional,
clean quantum antiferromagnets with a small, positive,
zero temperature ($T$)
stiffness $\rho_s$, but with the ratio $k_B T /  \rho_s $
arbitrary. Universal scaling forms for the uniform susceptibility
($\chi_u$),
correlation length($\xi$), and
NMR relaxation rate ($1/T_1$) are proposed and computed
in a $1/N$ expansion and by Mont\'{e}-Carlo simulations. For
large $k_B T/\rho_s$, $\chi_u (T)/T$ and $T\xi(T)$ asymptote to universal
values, while
$1/T_{1}(T)$ is nearly $T$-independent.
We find good quantitative agreement
with experiments and some numerical studies on $La_{2-\delta} Sr_{\delta} Cu
O_4$.
\end{abstract}
\narrowtext

The last few years have seen extensive theoretical and experimental
studies of two dimensional quantum Heisenberg antiferromagnets,
with particular attention to the antiferromagnetism in the cuprate
compounds~\cite{theory,numerics}.
On the theoretical side, most notable has been the work of
Chakravarty {\em et.al.\/}~\cite{CHN} who focused mainly on the low
temperature ($T$) properties of
systems with well established long-range N\'{e}el
order at $T=0$; their most detailed results were
in a regime in
which the fully renormalized, $T=0$, spin-stiffness $\rho_s$, was
not too small,
while the temperature satisfied $k_B T \ll  \rho_s$.
Under these conditions, the antiferromagnet could be treated as
a classical system, with all effects of quantum fluctuations being
absorbed into renormalization of the couplings.
At low $T$, there has been good agreement between
their results and experiments on $La_2 Cu O_4$~\cite{CHN}.
However, the experimental results at higher $T$ remain poorly
understood - there are clear deviations from the classical behavior and it is
expected that quantum fluctuations will play a more fundamental role.
Besides, in the lightly-doped cuprates, $\rho_{s}$ is likely to be quite
small, thus decreasing the $T$ range over which the renormalized-classical
behavior will hold. Finally,
  there are experimental realizations of frustrated two-dimensional
Heisenberg antiferromagnets~\cite{otherafm}, which, in all likelihood, have
a very small value of
$\rho_s$.

Our understanding of the experiments would clearly be improved by
precise theoretical predictions in low temperature regimes
other than $k_B T \ll  \rho_s$.
To this end, we discuss here some universal properties of clean
two-dimensional quantum Heisenberg antiferromagnets, with nearest-neighbor
exchange $J$, in which the
stiffness $\rho_s $ is `small', but non-zero. We will study the
physics when  $0< \rho_s \ll J$, $k_B T \ll J$, but the
ratio $k_B T/ \rho_s$ is allowed to be {\em arbitrary\/}.
The system is then controlled by
renormalization-group flows near the
the $T=0$ quantum
fixed-point separating
the N\'{e}el ordered
and quantum-disordered phases.
Our main new result will be that, in this regime, the absolute values
of the entire long-wavelength,
low-frequency, uniform and staggered spin susceptibilities are completely
universal functions of just three thermodynamic parameters: $\rho_s$, $c$,
and the ordered staggered moment $N_0$. The universal functions
depend {\em only} on the symmetry of the order parameter, and sensitivity to
all lattice-scale physics arises only through the values of $\rho_s$, $c$, and
$N_0$.
For small $k_B T/
\rho_s$, the $T$-dependence of our results
is similar to those already obtained in
Ref.~\cite{CHN}. For large $k_B T/ \rho_s$,
most of our results are new.
We will show that they are consistent
with the available experimental~\cite{Johnson,Yamada,Slichter}
and some of the numerical~\cite{Singh-Gelfand,Makivic,Sokol}
data on the uniform
susceptibility, correlation length and NMR relaxation rate for
undoped and weakly-doped $La_{2-\delta}Sr_{\delta}CuO_{4}$. We thus argue
 that the use of a `small' $\rho_s$ point-of-view
is not unreasonable even for the pure square lattice, spin-1/2,
Heisenberg antiferromagnet; while ordered at $T=0$, this system is
evidently close to the point where long-range-order vanishes.

Our results follow from some very general
properties of the $T=0$ quantum fixed-point separating the
magnetically-ordered and quantum-disordered phases. These properties are
expected to be valid in both undoped and doped antiferromagnets, though not
in the presence of randomness~\cite{Subir,random}. They are ({\em i\/}) the
fixed point is described by a continuum 2+1 dimensional field theory
which is `Lorentz'-invariant, and the spin-wave
velocity, $c$, remains non-singular through the phase
transition
({\em ii\/}) at $T=0$, on the
magnetically ordered side, there is a Josephson correlation length
$\xi_J$ which diverges at the quantum fixed-point;
near this fixed point $\rho_s$ equals $\hbar c \Theta/ \xi_J$ where $\Theta$ is
a
universal number~\cite{josephson,matt}; and
({\em iii\/}) Turning on a small $T$
places the critical field theory in a `slab' geometry which is
infinite in the two spatial directions, but of finite length $L_{\tau}
= \hbar c/(k_B T)$,
in the imaginary time ($\tau$) direction -
its consequences therefore follow from finite-size scaling.

\underline{Uniform susceptibility, $\chi_u$}: We first consider the
 response of the antiferromagnet to a static, spatially uniform,
external magnetic field (the extension to a field at finite wavevector $k$
or frequency $\omega$ will be omited here for brevity).
Such a field causes a uniform precession of all
the spins, which can be
removed by transforming to a rotating reference frame at the price of
a twist in the boundary conditions along the $\tau$
direction\cite{Fisher}. The response of the
system to this twisted boundary condition defines a stiffness, $\rho_{\tau}$,
which equals $\chi_u$.
However, the fixed point
is Lorentz invariant, and hence $\chi_{u}$
has the same scaling properties
as $\rho_s$.
Application of finite-size
scaling~\cite{binder}
then yields the following $T$ dependence for $\chi_u$
\begin{equation}
\chi_u ( T) = \left(\frac{g \mu_B}{\hbar c} \right)^2 k_B T ~\Omega_{Q}(x)
{}~~;~~~x \equiv
\frac{N k_B T}{ 2\pi \rho_s}
\label{chiuniv}
\end{equation}
where $g\mu_B/\hbar$ is the gyromagnetic ratio,
$N$ is the number of components of the order parameter, and $\Omega_Q (x)$ is
a {\em universal\/} function.
Note $x \propto \xi_J / L_{\tau}$, the length-ratio expected in finite-size
scaling
functions.
We have computed $\Omega_Q (x)$ in a $1/N$ expansion for
the $O(N)$ non-linear sigma model in 2+1 dimensions\cite{Polyakov}.
The $O(3)$ model describes the low-energy dynamics of
two-dimensional Heisenberg antiferromagnets on a square lattice.
The antiferromagnet also carries Berry phases, not present in the
$\sigma$-model, but these have been argued
to be irrelevant at the quantum fixed-point~\cite{Subir}. At $N= \infty$,
the scaling function $\Omega_Q (x)$ can easily be calculated:
\begin{displaymath}
 \Omega_Q^{N=\infty} (x) = \frac{1}{\pi x} +
\frac{\sqrt{4 + e^{-2/x}}}{2\pi e^{-1/x}} F(x)
\end{displaymath}
\begin{equation}
F(x) \equiv 2 ~\mbox{arcsinh} ( (1/2) e^{-1/x} )
\label{Ninfty}
\end{equation}

Of particular interest is the behavior of $\chi_{u}$ for large $x$.
The function $\Omega_Q^{N=\infty} (x)$ is analytic at $x=\infty$, and the
general principles of finite-size scaling~\cite{binder} suggest that
this remains true at finite $N$. Thus we expect that
$\Omega_Q ( x \rightarrow \infty ) = \Omega_{\infty} + \Omega_1 /x + \cdots$
with $\Omega_{\infty}$, $\Omega_1$ universal numbers. Combined with
(\ref{chiuniv}),
this implies that a plot of $\chi_u (T)$ vs. $T$ will be straight line at large
$T/\rho_s$ with universal slope and intercept, whose values are related to
$\Omega_{\infty}$ and $\Omega_1$ respectively.
At $N=\infty$ we obtain from (\ref{Ninfty}),
$\Omega_{\infty} = (\sqrt{5}/\pi) ~\mbox{ln}~[(\sqrt{5}+1)/2] \approx
0.3425$ and $\Omega_1 = 4\Omega_{\infty}/5$. We have computed
the first $1/N$ correction to $\Omega_{\infty}$ and indeed found that it
is a universal, regularization-independent number. We obtained
\begin{equation}
\Omega_Q ( x = \infty ) \equiv \Omega_{\infty} = 0.3425(1 - 0.63/N +
 \ldots).
\label{omegainf}
\end{equation}
We have also performed Mont\'{e}-Carlo simulations of a {\em classical\/}
$D=3$ Heisenberg ferromagnet on a cubic lattice, whose phase transition is
expected to be in the same universality class as the $O(3)$ sigma model.
We used a lattice of size $L \times L \times L_{\tau}$ ($L \leq 30$,
$L_{\tau} \leq 10$) at its known
critical coupling~\cite{holm} and computed $\rho_{\tau}$. It then follows
that $\Omega_{\infty} = \lim_{L_{\tau} \rightarrow \infty} \lim_{L
\rightarrow \infty} L_{\tau} \rho_{\tau}$ where the order of limits is
crucial. The result was
$\Omega_{\infty} = 0.25 \pm 0.04$,
in good agreement with the $1/N$ result at $N=3$.
Finally, there is an
analogy between $\Omega_{\infty}$ and another universal number
discussed recently - the universal conductivity, $\sigma_Q$, at the
superfluid-insulator transition~\cite{matt,sigq}.

We turn next to small $x$. The $N=\infty$ result (\ref{Ninfty})
gives the leading term $\Omega_Q (x\rightarrow 0) = 1/(\pi x)$ which
implies
$\chi_u (T=0) = 2 g^2 \mu_B^2 \rho_s /(\hbar^2 c^2 N)
\equiv (2/N) \chi_{\perp}$, where $\chi_{\perp}$ is the transverse
susceptibility.
This is in fact equal to the exact result expected from rotational averaging
of an ordered quantum $O(N)$ sigma model~\cite{CHN} - we have indeed found no
corrections in the $1/N$ expansion at $T=0$.
Furthermore, it was shown in Ref.~\cite{CHN} that there are no $T$-dependent
corrections to the isotropic $\chi_u$ in a classical lattice rotator model.
In contrast, for our quantum $O(N)$ sigma model, the $N=\infty$ result
contains a term linear in $T$ at small $T$. In the $1/N$ expansion
of this quantum model, the classical contributions appear as $\log{x}/N$ terms;
however, as expected, the coefficient of $\log{x}/N$
in $\chi_u$ was found to vanish.
These results imply
\begin{equation}
\Omega_Q ( x \rightarrow 0 ) = 1/(\pi x) + \Gamma_N  + \ldots
\end{equation}
where $\Gamma_N = 1/\pi + {\cal O}(1/N)$ is a universal number; some
$x$ dependence in $\Gamma_N$ at order $1/N^2$ is not ruled out.

\underline{Correlation Length, $\xi$:} The scaling dimension of $\xi$
is -1, and the finite-size scaling result for $\xi(T)$ is therefore:
\begin{equation}
\xi^{-1} (T) = (k_B T/(\hbar c)) X_Q (x),
\label{xiuniv}
\end{equation}
where $X_Q (x)$ is a universal function. As for $\chi_u$, there are no
non-universal factors on the right hand side. The numerical results
for $X_Q$ depend on the precise definition chosen for $\xi$: we use
$\xi^2 = -(1/S(k))(\partial S/\partial k^2) |_{k=0}$ where $S(k)$ is
the equal-time staggerd spin structure factor
(this definition is slightly different
from that used in Ref.~\cite{CHN}). At $N=\infty$, we found
\begin{equation}
X_Q^{N=\infty} (x) = F(x)\sqrt{2/[ 1
+ F(x)/ \sinh(F(x))]}
\label{xi}
\end{equation}

For large $x$, the properties of $\xi^{-1}$ are similar to those of $\chi_u$.
The function $X_Q (x)$ is expected to be analytic at $x=\infty$ with
$X_Q ( x \rightarrow \infty ) = X_{\infty} + X_1 /x + \ldots$; a
plot of $\xi^{-1} (T)$ vs. $T$ will be straight line at large
$T$ with a universal slope and intercept, whose values are related to
$X_{\infty}$ and $X_1$ respectively.
At $N=\infty$ we find $X_{\infty} = 0.998, X_1 = -0.990$.

For small $x$, the $N=\infty$ result (\ref{xi}) gives
$X_Q (x \rightarrow 0) = e^{-1/x}$. However, unlike $\chi_u$,
the renormalized-classical spin fluctuations make a strong contribution
to $\xi$, thus requiring careful consideration of the $\log{x}/N$ terms
in the $1/N$ expansion. We identified terms to order $(\log{x}/N)^2$,
exponentiated them and found
\begin{equation}
X_Q ( x \rightarrow 0) = Y_N x^{-1/(N-2)} e^{-N/((N-2)x)}
\label{xqsmallx}
\end{equation}
where $Y_N=$ is a universal constant, and $\lim_{N\rightarrow\infty} Y_N = 1$.
%We emphasize that both the factor
%$N/(N-2)$ in the exponent, and
%the $x$ dependence in the prefactor, result from the summation of the
%logarithmically divergent terms.
The analysis of
Chakravarty {\em
et. al.\/}~\cite{CHN}, valid for
$k_B T \ll  \rho_s $ but $\rho_s  /J$
arbitrary,
 obtained an identical functional form for $\xi (T)$, but
with $Y_N$ replaced by a non-universal, regularization dependent
prefactor whose determination for a square-lattice $2D$ $S=1/2$ Heisenberg
antiferromagnet required a lengthy calculation.
Our new result here is that, in the lowest order in $\rho_s /J$,
this prefactor is also universal, and can therefore be evaluated
in any convenient regularization scheme.

\underline{NMR relaxation rate, $1/T_1$:}
The relaxation of nuclear spins coupled to the antiferromagnetic order
parameter ({\em e.g.\/} $Cu$ nuclear spins in $La_2 Cu O_4$) is
given by $1/T_{1}(T) = \lim_{\omega \rightarrow 0} 2
\tilde{A}^{2}_{\pi} k_B T/(\hbar
\omega )
\int d^{2} k /(4\pi^2) \chi^{\prime\prime} ( k , \omega )$, where
$\chi^{\prime\prime} (k,\omega)$ is the imaginary part of the dynamic
staggered susceptibility of the underlying quantum antiferromagnet,
and $\tilde{A}_{\pi}$ is the bare
hyperfine coupling.
This determines the scaling dimension of
$1/T_1$ at the quantum fixed-point to be
$\eta$, the critical exponent associated with spin correlations at
criticality: $\eta = 8/(3 \pi^2 N) - 512/(27 \pi^4 N^2 ) + \ldots$
in a $1/N$ expansion~\cite{Ma} and the best current
value at $N=3$ is $\eta \approx 0.028$~\cite{holm}.  The
finite-size scaling form for
$1/T_1$ can be shown to be
\begin{equation}
1/T_1 (T) =  (2 \tilde{A}_{\pi}^2 N_0^2 / \rho_s ) x^{\eta} R_Q (x)
\label{nmruniv}
\end{equation}
where $R_{Q} (x)$ is a completely universal function.
Ref.~\cite{Chak-Orbach}
used the renormalized hyperfine coupling
$A_{\pi}$ which is
$A_{\pi} = \tilde{A}_{\pi} N_0$. Note the complete absence of non-universal
normalization factors in (\ref{nmruniv}).

We now consider the limiting behavior of $R_Q (x)$ for large and small $x$.
As before, at large $x$,
$R_Q ( x \rightarrow \infty ) =
R_{\infty}$,  a positive constant;
the small value of $\eta$ then implies that
$1/T_1$ is {\it{essentially $T$ independent}} at high $T$.
To leading order in $1/N$,  $R_{\infty}$ can be deduced from the results of
Ref.~\cite{Subir}
to be $R_{\infty} = 0.06/N$. Note the factor of $1/N$ -
$\chi^{\prime\prime} (k, \omega)$ is finite at $\omega \rightarrow 0$  only
due to the self-energy corrections. At small $x$, dynamical
scaling\cite{Halperin}
predicts that  $R_{Q}(x) \propto
\xi(x)$. The $1/N$ expansion is again singular when $x\ll 1$
and we will not
discuss it here.

\underline{Comparison with numerical and experimental results:}
We have so far presented general scaling forms for the magnetic properties
of a two-dimensional quantum antiferromagnet which has
$\rho_{s} \ll J$. Explicit scaling functions can be calculated at $N=\infty$,
and examination of $1/N$ corrections has been limited to those for
$\Omega_{\infty}$. These corrections were however quite small, and
we expect, in general, that $1/N$ expansion is robust and numerically
quite accurate for large values of $x$.
On the other hand, at small
$x$, the $1/N$ expansion is logarithmically singular, and eventually changes
the leading singularity in some of the scaling functions at $x=0$;
the final low-$T$ behavior is the same as
that in the renormalized-classical
scaling theory of Ref\cite{CHN}.
The crossover between small and large $x$ should occur for $x$ around
unity.
Thus for $x\geq 1$ (but such that long-wavelength description is
still valid), it is quite likely that $1/N$ expansion
will describe the experimental data better than the renormalized-classical
theory, which,
strictly speaking, is  valid only for
$x \ll 1$. In a square lattice, nearest-neighbor, $S=1/2$
Heisenberg antiferromagnet, $2\pi \rho_s  \approx
 1.13 J$ (Ref\cite{Oguchi}) and we therefore expect that our large-$N$, large
$x$
results should work for $x\geq 1$, i.e., for
$k_B T \geq 0.35 J$.

The absence of any renormalized-classical corrections to $\chi_u$ makes
it an ideal candidate for testing our theory; the $1/N$ expansion should
become accurate even at fairly small values of $x$.
We start
with the numerical results for $\chi_u (T)$ on the square
lattice $S=1/2$ antiferromagnet. There have been high-$T$ series
expansions~\cite{Singh-Gelfand}, quantum
Mont\'{e}-Carlo~\cite{Makivic} and finite cluster
calculations~\cite{Sokol}.
Their results all show that $\chi_u (T)$
obeys a Curie-Weiss law at high T, reaches a maximum at
$k_B T \sim J$ and then falls to a finite value at $T=0$
 which is rather close to the rotationally averaged $1/S$ result
$(\hbar/g\mu_B)^{2} \chi_u (T=0)
\approx 0.04/Ja^2$, where $a$ is the lattice spacing.
%The
%initial $T$ dependence of $\chi_u$ ($T<0.3J$) is clearly
%sublinear although the data are not accurate enough to verify the
%$T^{3}$ dependence.
For $0.35J<T<0.55J$,
both series expansions~\cite{Singh-Gelfand} and
Monte-Carlo~\cite{Makivic} calculations report a {\it{linear}}
$T$ dependence of $\chi_u (T)$ (Fig.\ref{Fig.1}).
Also plotted is our theoretical prediction
of Eqns (\ref{Ninfty},\ref{omegainf}) which, over the range of
$x$ values used in the figure, is well approximated by
$(\hbar/g\mu_B )^2 (Ja^{2} \chi_{u}(T)) \approx 0.037 x (1 + \alpha /x)
$, $\alpha = 0.8 + {\cal O}(1/N)$. This is remarkably close to the
best fit to the data of Ref.~\cite{Makivic} which
gives $0.037 x (1 + 0.775/x)$.

We consider next measurements of $\chi_u (T)$ in weakly doped
$La_{2-\delta}Sr_{\delta}CuO_{4}$. The interpretation of the experimental data
 even above the zero-doping $T_{N}$ requires caution
because one has to subtract
Van-Vleck, core and diamagnetic contributions from the measured $\chi_u (T)$.
Besides, at nonzero doping, a Pauli-like contribution from valence fermions
should also be subtracted. Nevertheless, after subtraction was carried out,
it was found~\cite{Johnson,Andy} that
there is a large amount of universality in the susceptibility data
at different
doping concentrations  and
 the measured susceptibility is clearly {\it{linear}} in $T$
in the temperature interval $400-700 K$ (i.e., $0.3J - 0.5J$ for $J \sim
0.12$ev). The measured slope of
$(\hbar/g\mu_B )^2 (Ja^{2} \chi_{u}(T))$ vs. $x$
is about $0.04$ which is very close to our result ($0.037$).

Now the $^{63}Cu$ spin-lattice
relaxation rate, $1/T_{1}$, in $La_2 Cu O_4$. At very
low $T$ ($x$ small),
the theory of Ref.~\cite{Chak-Orbach}
predicts that
$1/T_{1} \propto e^{-3/x}$; this
is consistent with recent observations~\cite{Slichter}. However,
the present theory  predicts that $1/T_1$ becomes nearly $T$ independent for
$x > 1$ or $T >  0.35J$.
This has in fact already been observed
in series expansions\cite{Singh-Gelfand} and finite cluster
calculations\cite{Sokol} for the square lattice
antiferromagnet. More importantly, a flattening in $1/T_{1}(T)$
 has recently
been observed in the experiments on $La_{2}CuO_{4}$ \cite{Slichter}. We
calculated, from our results above,
the limiting large $T$ value of $1/T_{1}$
for the same values of parameters as
were used in the low-$T$ fit~\cite{Slichter}
and found  $1/T_1 \approx 3.3\times 10^3 sec^{-1}$;
this is in good agreement with the experimental result
$1/T_{1} \approx 2.7\times 10^3 sec^{-1}$.  Furthermore,
the experimental
$T$ range over which $1/T_1$ is nearly $T$ independent increases upon doping.
This is also consistent with our results because $\rho_{s}$ is expected to
decrease with doping, thus pushing the system into larger $x$ for the same $T$.

The agreement between our theory and the data
for $\xi$ is however not so
good. Detailed measurements of $\xi$
 in $La_2 Cu O_4$ have been performed at low $T$, where the system is
in the renormalized-classical region\cite{Yamada}.
At the highest experimentally accessible
$T$ ($=560$K  for $J=1460$K), our result $\xi^{-1} = 0.023 \mbox{\AA}^{-1}$ is
not
far from the experimental value of $\xi^{-1} = 0.03 \mbox{\AA}^{-1}$.
We also compared our results
with the numerical data for $\xi$ at higher $T$\cite{Makivic,Manous-MC}.
For $T > 0.35 J$ this data obeys quite well
$\xi^{-1} \propto x (1 - \gamma/x)$ where $\gamma$ is close to one. However,
the overall factor in $\xi$ in the fit is close to twice our $N=\infty$ result.
This discrepancy is probably due to the fact that the strong singular
corrections
in $X_Q (x)$ at small $x$, cause the
crossover from small to large $x$ behavior to occur at a larger $x$.
Note however that the Mont\'{e}-Carlo calculations in the quantum-critical
region\cite{Man} yield the value of
$X_{\infty} = 1.25$, which is close to
our $N=\infty$ result of $X_{\infty} = 0.998$.

To conclude, we have considered in this paper the magnetic properties of
two-dimensional
quantum antiferromagnets. We argued that there exists a $T$
range where the classical low-$T$ description is no longer valid and the
behavior
of observables is governed by the renormalization-group flows near the $T=0$
quantum fixed-point. A comparison with the experimental data for the uniform
susceptibility and  $^{63}Cu$ spin-lattice relaxation rate
shows that the intermediate behavior has been observed in the range
$0.35J <T< 0.55J$ in $La_2 Cu O_4$.

We thank N. Read and  A. Sokol
for discussions. This work has been
supported by NSF Grant DMR88-57228.

\figure{`Experimental'~\cite{Makivic} (squares) and theoretical (line)
results
for the uniform susceptibility $\overline{\chi}_u = (3 J (a\hbar/g \mu_B)^2)
\chi_u$ of a square lattice spin-1/2 Heisenberg antiferromagnet ($a$ is the
lattice spacing).
There are {\em no} adjustable parameters in the theoretical result
(\ref{chiuniv}).
Over the range of $x$ plotted, the function $\Omega_Q (x)$ is very close
to its large $x$ behavior
$\Omega_Q (x) \approx \Omega_{\infty} (1 + 0.8/x)$. We used this large $x$
result
with
$\Omega_{\infty}$ from (\ref{omegainf})
at $N=3$.
The theoretical and experimental slopes agree remarkably well. The good
agreement
in the intercept is somewhat surprising as
 its theoretical value ($=0.8$) is known only
at $N=\infty$.
\label{Fig.1}}


\begin{references}
\bibitem{random} Randomness is expected to be a relevant perturbation
at the $T=0$ quantum fixed-point~\cite{Subir}.
Nevertheless, it is probably innocuous at the
rather high $T$'s at which experimental comparisons are made in this paper.
\bibitem{sigq} At the quantum fixed-point,
we have $\Omega_{\infty} \sim
\chi_u (T)/T \sim \rho_{\tau} (T, \omega=0) / T$,
while $\sigma_Q \sim \rho_s (T=0, \omega)/(i \omega)$, where $\omega$
is the measurement frequency. The two quantities are thus related by
a Lorentz rotation and the interchange of $\omega$ and $T$. However,
computation of
$\Omega_{\infty}$ requires use of finite temperatures, which makes
the analysis considerably more difficult than that for
$\sigma_Q$.
\end{references}
\end{document}